\documentclass[referee]{raa}
\usepackage{graphicx,times}
\usepackage{natbib}
\usepackage{amssymb,amsmath}
\bibpunct{(}{)}{;}{a}{}{,}

\usepackage[a4paper=true,dvipdfm=true,pagebackref=true]{hyperref}
\hypersetup{colorlinks = true, linkcolor = green, anchorcolor =
red, citecolor = blue, filecolor = red, pagecolor = red, urlcolor
= red}

\begin{document}

   \title{NSVS 2569022: a peculiar binary among the W~UMa stars with extremely small mass ratios
}

   \volnopage{Vol.0 (20xx) No.0, 000--000}      
   \setcounter{page}{1}          

   \author{Diana P. Kjurkchieva
      \inst{1}
   \and Velimir A. Popov
      \inst{1}
   \and Nikola I. Petrov
      \inst{2}
   }

   \institute{Department of Physics and Astronomy, Shumen University, 115 Universitetska Str., 9712 Shumen,
   Bulgaria; {\it d.kyurkchieva@shu.bg}\\
        \and
   Institute of Astronomy and NAO, Bulgarian Academy of Sciences, 72 Tsarigradsko Shose Blvd., 1784 Sofia, Bulgaria\\
\vs\no
   {\small Received~~20xx month day; accepted~~20xx~~month day}}

\abstract{Photometric observations of the W UMa binary NSVS
2569022 are presented. The light curve solution reveals that both
components are of F spectral type (temperatures $T_1$=$T_2$=6100
K). NSVS 254037 undergoes total eclipse of W subtype and the mass
ratio is well-determined. Its exclusive small value of only 0.077
implies that the target probably goes to instability and possible
merger. This value ranks NSVS 2569022 in sixth place among the
binaries with smallest mass ratio. On the base of empirical
relation ''period -- total mass'' for low mass-ratio binaries we
estimated the global parameters of NSVS 2569022: masses $M_1$=1.17
M$_{\odot}$ and $M_2$=0.09 M$_{\odot}$; radii $R_1$=1.19
R$_{\odot}$ and $R_2$=0.38 R$_{\odot}$; luminosities $L_1$=1.73
L$_{\odot}$ and $L_2$=0.17 L$_{\odot}$. An analysis of the
characteristics of binaries with extremely low-mass ratios is
made. NSVS 2569022 turns out peculiar binary among the W UMa stars
with extremely small mass ratios due to its unexpected small
fill-out factor of only 0.014 (slightly overcontact
configuration). \keywords{binaries: eclipsing; binaries: close;
stars: fundamental parameters; stars: individual (NSVS 2569022)} }

   \authorrunning{D. Kjurkchieva, V. Popov \& N. Petrov}            
   \titlerunning{NSVS 2569022: a peculiar binary among the W~UMa stars with extremely small mass ratios}  

   \maketitle

%
%

\section{Introduction}

The components of W UMa systems lie between the inner and outer
Lagrangian equipotential surfaces. Because of the similarity of
effective temperatures of their components, the eclipse depths are
practically independent on the temperatures, but depend strongly
on the mass ratio $q$ as well as on geometrical parameters orbital
inclination $i$ and the degree of contact $f$. The case is
different for detached binaries where the differences in component
temperatures have a significant effect on the eclipse depths
(\citealt{Rucinski+01}).

The study of W UMa stars is essential for the modern astrophysics
because they are probes for investigation of the processes of
tidal interactions, mass loss and mass transfer, angular momentum
loss (\citealt{Martin+etal+11}). However, the mechanism of energy
transfer, the W phenomenon (the hotter component is the smaller
star) and the future fate (tight binaries or merger), are still
debatable problems of the W UMa stars. Their solution requires a
knowledge of the fundamental parameters of a number of such
binaries.

The mass ratio is an important parameter for each binary star,
particularly it determines the W/A subtype of the W UMa systems
(\citealt{Binnendijk+70}). But the mass ratio is poorly determined
on photometric data of those binaries which undergo partial
eclipses (\citealt{Rucinski+01, Terrell+Wilson+05}). On the other
hand spectral data of W UMa stars are rarely available due to
their faintness (\citealt{Rucinski+02}). Moreover, even when the W
UMa stars have spectral observations, their spectral mass ratios
are not precise due to the highly broadened and blended spectral
lines (\citealt{Frasca+00, Bilir+etal+05,
Dall+Schmidtobreick+05}).

Firstly \cite{Webbink+76}  established that there is a cut-off of
mass ratio: for $q < q_{min}$ the binary system quickly merges
into a single, rapidly rotating star. The instability occurs when
$J_{orb} \approx 3J_{rot}$. \cite{Rasio+95} theoretically
calculated $q_{min} \simeq$ 0.09 for a contact binary containing
two unevolved MS stars whose primary is mostly radiative (the
value is bigger for convective envelope). \cite{Li+Zhang+06} found
$q_{min} \sim $ 0.071--0.076 assuming that the W UMa systems
rigorously comply with the Roche geometry. For overcontact
binaries \cite{Arbutina+07} determined theoretically $q_{min}
\sim$ 0.094--0.109. Later, by including the effects of higher
central condensation due to rotation, \cite{Arbutina+12} found
$q_{min}$ = 0.070--0.074 for overcontact binaries with fill-out
factor $f$ = 0 -- 1. \cite{Jiang+etal+10} argued that $q_{min}$
depends on the primary mass and structure and can reach up to
0.05. \cite{Yang+Qian+15} obtained even $q_{min}$ = 0.044. It was
supposed that the lower limit of mass ratio depends on the
fill-out factor $f$ (\citealt{Rasio+95, Rasio+Shapiro+95}) but
this dependence was not studied. The majority of the foregoing
theoretical investigations has been invoked by the discoveries of
several binaries with extremely low mass ratios ($q_{min}< 0.09$,
see further Table 3).

Observations of low mass ratio systems is valuable to understand
the dynamical evolution of binaries and the formation of blue
stragglers and FK Com-type stars
(\citealt{Eggleton+KiselevaEggleton+01}). In this paper we present
photometric observations of the eclipsing binary NSVS 2569022 (RA
= 10$^{h}$10$^{m}$42$^{s}$.72, Dec = +67$^{\circ}$39$'$32$''$.4, V
= 10.1 mag) whose modeling shows that it is the sixth target among
the W UMa systems with the smallest values. \cite{Gettel+etal+06}
have classified NSVS 2569022 as EW type with period $P$ =
0.2877891 d and amplitude 0.21 mag.

\section{Observations}

The CCD photometric observations of the target in Sloan $g', i'$
bands were carried out on Feb 13 and 14 2015 at Rozhen Observatory
with the 30-cm Ritchey Chretien Astrograph (located into the
\emph{IRIDA South} dome) using CCD camera ATIK 4000M (2048
$\times$ 2048 pixels, 7.4 $\mu$m/pixel, field of view 35 x 35
arcmin). The exposures in $g'$ and $i'$ were 30 and 60 s, the mean
errors in the two filters were $\sim$0.002 mag.

The standard procedure was used for the reduction of the
photometric data (de-biasing, dark frame subtraction and
flat-fielding) by software \textsc{AIP4WIN2.0}
(\citealt{Berry+Burnell+05}).

The light variability of the target was estimated with respect to
nearby comparison (constant) stars in the observed field (ensemble
photometry). A check star served to determine the observational
accuracy and to check constancy of the comparison stars (Table 1).
We performed the ensemble aperture photometry with the software
\textsc{VPHOT}. The magnitudes of the comparison and check stars
were taken from the catalogue APASS DR9
(\citealt{Henden+etal+16}). The transformation of the obtained
instrumental magnitudes to standard ones was made as in
\cite{Kjurkchieva+etal+17}.

\begin{table*}
\scriptsize
\begin{center}
\caption[] {Magnitudes of standards \label{t1}}
\begin{tabular}{ccrrrr}
\hline\hline
Label   & Star ID     & RA          & Dec           & $g'$   & $i'$  \\
\hline
Target & NSVS 2569022 & 10 10 42.72 &  +67 39 32.40 & 10.357 & 9.784 \\
Chk & UCAC4-789-020996& 10 11 02.25 &  +67 37 17.16 & 13.412 & 12.301 \\
C1 & UCAC4-789-021025 & 10 13 01.97 &  +67 45 25.22 & 11.525 & 10.859 \\
C2 & UCAC4-790-020961 & 10 08 55.89 &  +67 48 49.73 & 12.668 & 11.473 \\
C3 & UCAC4-789-021011 & 10 12 04.17 &  +67 40 36.75 & 12.687 & 11.161 \\
C4 & UCAC4-789-021009 & 10 11 57.27 &  +67 39 46.56 & 12.270 & 11.604 \\
C5 & UCAC4-788-021201 & 10 11 15.26 &  +67 35 07.06 & 13.006 & 11.907 \\
C6 & UCAC4-788-021202 & 10 11 18.72 &  +67 32 27.09 & 12.065 & 11.508 \\
C7 & UCAC4-788-021204 & 10 11 48.18 &  +67 28 03.79 & 11.987 & 10.812 \\
\hline
\end{tabular}
\end{center}
\end{table*}

Low resolution spectrum (10.84 \AA/px) of the target was obtained
on Feb 13 2015 by IRIDA-South instrumentation using SA-200 grating
with 200 groves/mm (Fig. 1). It was calibrated by \textsc{RSpec}
software (\citealt{Field+11}) according to the instrumental
response of the setup.

\begin{figure}
\centerline{\includegraphics[width=8cm]{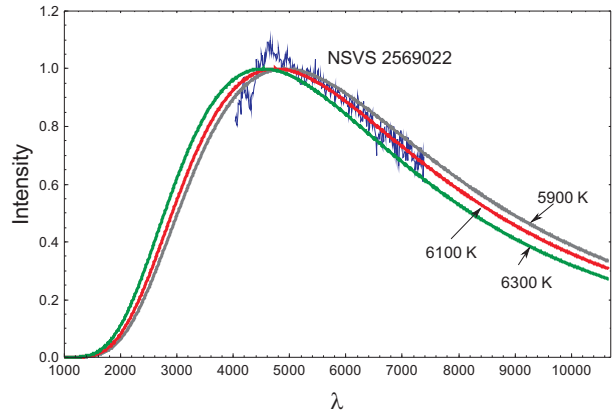}}
\caption[]{Low-resolution spectrum of NSVS 2569022 \label{Fig1}}
\end{figure}

\section{Modeling}

The photometric data were modelled using the \textsc{PHOEBE} code
(\citealt{Prsa+Zwitter+05, Prsa+etal+11, Prsa+etal+16}) which is
based on the Wilson--Devinney (WD) code
(\citealt{Wilson+Devinney+71, Wilson+79, Wilson+93}). It allows
simultaneous modeling of photometric data in a number of filters
and provides a graphical user interface.

\begin{figure}
\centerline{\includegraphics[width=8cm]{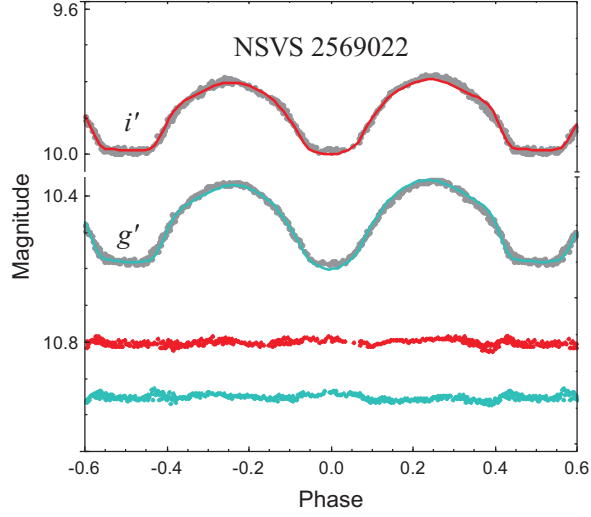}} \caption{Top:
folded light curves of NSVS 2569022 and their fits; Bottom:
corresponding residuals (shifted vertically by different amount to
save space); \label{Fig2}}
\end{figure}

We determined in advance the temperature of NSVS 2569022 as
$T_{m}$=6100$\pm$50 K by comparison of its low-resolution spectrum
with sample of Planck curves (Fig. 1). There are two deviations of
the observed spectrum from the (normalized) Planck curves:
protruding peak and steeper left branch (the same deviations might
be seen in the Sun spectrum). The obtained spectral temperature of
6100 K is bigger than the photometric values (Table 2) determined
by different color indices. The possible reason could be the using
of data without de-reddening. For instance, the temperature
corresponding to our original data is 5870 K (Table 2) while the
procedure of their de-reddening led to $g'-i'$ = 0.499
(\citealt{Schlafly+Finkbeiner+11}) and correspondingly to
temperature of 6030 K, very close to the spectral value of 6100 K.

\begin{table*}
\scriptsize
\begin{center}
\caption[]{Target temperature from photometric observations
(without de-reddening)\label{t2}}
\begin{tabular}{ccccc}
\hline\hline
Index   & Value     & $T$     & Color-T relation  & Source    \\                                               \\
\hline
$J-K$    &   0.417   & 5650    & \cite{Cox+00}            & \cite{Cutri+etal+03}  \\
$B_T-V_T$, $B_T-J$, $V_T-H$, $V_T-K$, $J-K$ &  & 5932 & \cite{Valenti+Fischer+05} & \cite{Ammons+etal+06} \\
$B-V$     &   0.560   & 5800    & \cite{Worthey+Lee+11}   & \cite{Henden+etal+16}     \\
$B-V$     &   0.576   & 5760    & \cite{Worthey+Lee+11}   & \cite{Terrell+etal+12}    \\
$V-Ic$    &   0.693   & 5750    & \cite{Worthey+Lee+11}   & \cite{Terrell+etal+12}      \\
$V-Ic$    &   0.799   & 5460    & \cite{Worthey+Lee+11}   & \cite{Droege+etal+06}  \\
$g'-i'$   &   0.573   & 5870    & \cite{Covey+etal+07}    & This work             \\
\hline
\end{tabular}
\end{center}
\end{table*}

Fixed coefficients of gravity brightening $g_1 = g_2$ = 0.32 and
reflection effect $A_1 = A_2$ = 0.5 appropriate for stars with
convective equilibrium were assumed. Initially, we used a linear
limb-darkening law whose coefficients for each component and each
color were updated according to the tables of van Hamme (1993).

Since the light curve shapes (Fig. 2) implied overcontact
configuration, we employed the mode ''Overcontact binary not in
thermal contact''.

To search for a fit we fixed $T_{1}$ = $T_{m}$ and varied
simultaneously the initial epoch $T_{0}$, period \emph{P} (around
its known value), secondary temperature $T_{2}$ (around $T_m$),
orbital inclination $i$, mass ratio $q$ and potential $\Omega$.
The data in two bands were modelled simultaneously.

In order to reproduce the light curve distortion we added cool
spot whose parameters (longitude $\lambda$, latitude $\beta$,
angular size $\alpha$ and temperature factor $\kappa$) were
adjusted.

After reaching the best fits, we carried out also solutions for
quadratic and logarithmic limb-darkening laws but they did not
lead to better fits. So, further we present the results
corresponding to a linear limb-darkening law.

The final values of the fitted parameters are: $T_0$ =
2457068.471732(88); $P$ = 0.287797(4) d; $\Omega$ = 1.88729(5);
$q$ = 0.0777(3); $i$ = 76$^{\circ}$.29(8); $T_2$ = 6100(25) K. The
synthetic light curves corresponding to our solution are shown in
Fig.~2 as continuous lines while Figure 3 exhibits the target 3D
configuration.

\begin{figure}
\centerline{\includegraphics[width=6cm]{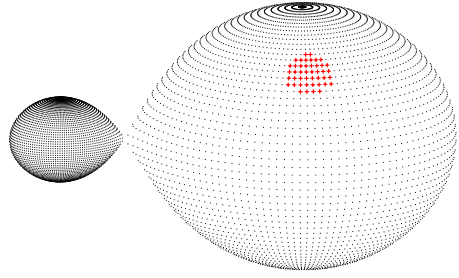}}
\caption{3D configuration of NSVS 2569022 \label{Fig3}}
\end{figure}


\textsc{PHOEBE} yields as output parameters the potentials
$\Omega(L_1)$ and $\Omega(L_2)$. They allowed us to calculate the
fill-out factor as $f = [\Omega - \Omega(L_1)]/[\Omega(L_2) -
\Omega(L_1)]$ = 0.014.

To estimate the global parameters of the target we used the
empirical relation $M_{tot} = 0.5747 + 2.3734 \times P$ of Yang
$\&$ Qian (2015) derived from sample of 46 low mass-ratio binaries
with $q\leq$ 0.25. We calculated $M_{tot}$ of NSVS 2569022 as
1.258 M$_{\odot}$. It makes possible calculation of the global
parameters : orbital axis $a$ = 1.98 R$_{\odot}$; masses $M_1$ =
1.168(13) M$_{\odot}$ and $M_2$ = 0.090(2) M$_{\odot}$; radii
$R_1$ = 1.188(8) R$_{\odot}$ and $R_2$ = 0.378(6) R$_{\odot}$;
luminosities $L_1$ = 1.727(13) L$_{\odot}$ and $L_2$ = 0.175(9)
L$_{\odot}$.

\section{Analysis of the results}

The main results from the light curve solution are as follows.

(1) We determined initial epoch of the target. The known value
of its orbital period phased well our data.

(2) The stellar components are of F spectral type.

(3) The target undergoes total eclipse.

(4) NSVS 2569022 has slightly overcontact configuration with
fill-out factor 0.014 (Fig. 3).

(5) The target is of W subtype.

(6) The different light levels at the two quadratures were
reproduced by cool spot on primary component with angular size of
10$^{\circ}$ and temperature factor of 0.9.

(7) The solution of NSVS 2569022 turned out low-sensitive to the
component temperatures and orbital inclination $i$ but very
sensitive to the mass ratio $q$ and potential $\Omega$. The
attempts to obtain fit with $\Omega$ close to $\Omega_2$
corresponding to big fill-out factor $f$, were not successful.

(8) NSVS 2569022 has an extremely small mass ratio. It was
determined photometrically on the base of fitting of wide total
eclipse and may be adopted with a big confidence. This conclusion
is supported by the coincidence of the values of photometric
($q_{ph}$) and spectral ($q_{sp}$) mass ratios for
totally-eclipsed binaries established by the investigation of
sample of low mass-ratio ($q\leq$ 0.25) systems (Table 1 in
\cite{Yang+Qian+15}).

(9) The values of stellar masses of NSVS 2569022 support
the conclusion of \cite{Arbutina+12} that the extreme mass ratio W
UMa systems (with $q \leq$ 0.1) represent an interesting class of
objects in which "normal", approximately one solar mass
main-sequence star is in contact with a significantly less massive
companion, $M_2 \sim$ 0.1 M$\odot$.

(10) According to the empirical relation
$J_{rot}/J_{orb}=0.5104-3.7738 \times q+8.2817 \times q^2$ for the
low mass-ratio overcontact binaries (\citealt{Yang+Qian+15}), NSVS
2569022 should have ratio $J_{rot}/J_{orb}$=0.269. This value is
close but smaller than the cut-off of 0.33.



\begin{table*}
\scriptsize
\begin{center}
\caption[]{List of known binaries with $q <$ 0.082. The sign
* means photometric mass ratio (for total eclipses),
the sign + means spectral mass ratio. \label{t3}}
\begin{tabular}{ccrrrrr}
\hline\hline
Target               & $q$   & $f$  & $P$ (d)     & $dP/dt$ (d yr$^{–1}$)  & $T$  & Reference  \\
\hline
V53 (M4)             & 0.060*&      & 0.308448705 &                        & 9800 & \cite{Kaluzny+etal+13} \\
V857 Her             & 0.065*& 0.84 & 0.38222952  & +2.90$\times$10$^{-7}$ & 8300 & \cite{Qian+etal+05} \\
SX Crv               & 0.066+&      & 0.3166      &                        & 6200 & \cite{Rucinski+etal+01}  \\
                     &0.0715+& 0.27 & 0.3165992   & -1.67$\times$10$^{-7}$ & 6300 & \cite{Zola+etal+04} \\
ASAS J083241+2332.4  & 0.068*& 0.5  & 0 311329    & +8.854$\times$10$^{-7}$& 6500 & \cite{Sriram+etal+16}  \\
AW UMa               & 0.075+&      &             &                        & 6850 & \cite{Rucinski+92} \\
                     & 0.080*& 0.84 & 0.43873     & -1.42$\times$10$^{-5}$ & 7000 & \cite{Pribulla+etal+09}  \\
NSVS 2569022         & 0.077*& 0.014& 0.2877891   &                        & 6100 & this paper\\
V870 Ara             & 0.082 & 0.96 & 0.399722    &                        & 6000 & \cite{Szalai+etal+07} \\
\hline
\end{tabular}
\end{center}
\end{table*}

\begin{table}
\scriptsize
\begin{center}
\caption[]{Eclipse times
 \label{t4}}
 \begin{tabular}{ccc}
 \hline\hline
type& HJD time                   & source  \\
\hline
I  & 2457067.32025 $\pm$ 0.00014 & this work\\
II & 2457068.32861 $\pm$ 0.00014 & this work\\
II & 2457068.61689 $\pm$ 0.00014 & this work \\
II & 2457449.66042 $\pm$ 0.00018 & AAVSO \\
II & 2457461.74881 $\pm$ 0.00017 & AAVSO \\
I  & 2457473.69294 $\pm$ 0.00042 & AAVSO \\
\hline
 \end{tabular}
\end{center}
\end{table}


\begin{table*}
\scriptsize
\begin{center}
\caption[]{List of binaries with $q <$ 0.19. The sign
* means only photometric mass ratio (for total eclipses),
the sign + means only spectral mass ratio.
\label{t5}}
\begin{tabular}{ccrrrcc}
\hline\hline
Target               & $q$   & $f$  & $P$ (d)     & $dP/dt$ (d yr$^{-1}$)  & $T$  & Reference  \\
\hline
AW CrB               & 0.101*& 0.75 & 0.3609      & +3.58$\times$10$^{-7}$ & 6100 & \cite{Broens+13} \\
DN Boo               & 0.103 & 0.64 & 0.4476      &                        & 6100 & \cite{Senavci+etal+08}  \\
ASASJ082243+1927     & 0.106*& 0.72 & 0.28        &                        & 6000 & \cite{Kandulapati+etal+15} \\
V1911 Cyg            & 0.107+& 0.68 & 0.313377    & +4.5$\times$10$^{-7}$  & 6500 & \cite{Rucinski+etal+08, Zhu+etal+11}   \\
CK Boo               & 0.11  & 0.72 & 0.3551534   & +9.79$\times$10$^{-8}$ & 6400 & \cite{Yang+etal+12}  \\
FG Hya               & 0.112 & 0.85 & 0.3278      & -1.96$\times$10$^{-7}$ & 6000 & \cite{Lu+Rucinski+99, Qian+Yang+05} \\
GR Vir               & 0.12+ & 0.78 & 0.3469788   & -4.32$\times$10$^{-7}$ & 6200 & \cite{Rucinski+Lu+99, Qian+Yang+04}   \\
DZ Psc               & 0.13  & 0.897& 0.3661278   & +7.43$\times$10$^{-7}$ & 6200 & \cite{Yang+etal+13}   \\
V776 Cas             & 0.138 & 0.77 & 0.4404      &                        & 6700 & \cite{Rucinski+etal+01, Zola+etal+04} \\
V345 Gem             & 0.142*& 0.73 & 0.2747736   & +5.88$\times$10$^{-8}$ & 6200 & \cite{Yang+etal+09}    \\
V710 Mon             & 0.143*& 0.60 & 0.4052      & +1.95$\times$10$^{-7}$ & 6000 & \cite{Yang+etal+09}    \\
V410 Aur      & 0.143--0.183 & 0.52 & 0.36634699  & +8.22$\times$10$^{-7}$ & 5900 & \cite{Yang+etal+05}   \\
HV Aqr               & 0.145*& 0.56 & 0.3734      & -8.84$\times$10$^{-8}$ & 6500 & \cite{Li+Qian+13}   \\
XY LMi               & 0.148*& 0.74 & 0.43688773  & -1.67$\times$10$^{-7}$ & 6100 & \cite{Qian+etal+11}   \\
EM Psc               & 0.149*& 0.95 & 0.34395922  & +3.97$\times$10$^{-6}$ & 5100 & \cite{Qian+etal+08}   \\
ASAS J113031–0101.9  & 0.15  & 0.50 & 0.270969    &                        & 6000 & \cite{Pribulla+etal+09}   \\
TYC 3836-0854-1      & 0.15  & 0.76 & 0.3961      &                        & 6000 & \cite{Acerbi+etal+14}   \\
V728 Her             & 0.16  & 0.81 & 0.44625     & +1.92$\times$10$^{-7}$ & 6600 & \cite{Erkan+Ulas+16}   \\
AH Aur               & 0.165*& 0.75 & 0.4941      &                        & 6300 & \cite{Gazeas+etal+05}  \\
TV Mus               & 0.166*& 0.74 & 0.4456794   &                        & 5900 & \cite{Qian+etal+05}   \\
AH Cnc               & 0.168*& 0.58 & 0.3604412   & +3.99$\times$10$^{-7}$ & 6300 & \cite{Qian+etal+06}     \\
IK Per               & 0.17* & 0.60 & 0.67603467  & -2.5$\times$10$^{-7}$  & 8600 & \cite{Zhu+etal+05}  \\
TYC 1337-1137-1      & 0.172*& 0.76 & 0.475511    &                        & 7800 & \cite{Liao+etal+17}   \\
CU Tau               & 0.177*& 0.50 & 0.41341521  & -1.81$\times$10$^{-6}$ & 5900 & \cite{Qian+etal+05}  \\
V728 Her             & 0.179 & 0.71 & 0.4713      & -1.81$\times$10$^{-6}$ & 6700 & \cite{Nelson+etal+95}  \\
Y Sex                & 0.180 & 0.64 & 0.4198      &                        & 6100 & \cite{McLean+Hilditch+83, Yang+Liu+03}  \\
MW Pav               & 0.182 & 0.51 & 0.7950      &                        & 6700 & \cite{Lapasset+80, Rucinski+Duerbeck+06} \\
XY Boo               & 0.185 & 0.56 & 0.3705524   & +6.25$\times$10$^{-7}$ & 6300 & \cite{Yang+etal+05} \\
V2388 Oph            & 0.186 & 0.77 & 0.8023      &                        & 6900 & \cite{Rucinski+etal+02, Yakut+etal+04} \\
TYC 3836-0854-1      & 0.19* & 0.79 & 0.415547    &                        & 7800 & \cite{Acerbi+etal+14, Liao+etal+17} \\
HV UMa               & 0.19  & 0.62 & 0.7108      &                        & 7200 & \cite{Csak+etal+00}  \\
\hline
 \end{tabular}
\end{center}
\end{table*}


\section{Extreme low mass ratio}

The most important peculiarity of NSVS 2569022 is its extremely
low mass ratio of 0.077. This result implies that the target
probably goes to instability and possible merger.

So far only 6 binaries with $q <$ 0.082 were known (Table 3).
Three of them are with mass ratios $q_{ph}$ determined
photometrically, from their total eclipses (as our case). One
should take into account that the determination of $q_{sp}$ of
extremely small mass-ratio binaries is exclusive difficult: on the
one hand the primary radial velocity is negligible, on the other
hand the secondary component is faint and invisible in the
spectra. Hence, an important possibility to determine global
parameters of low mass-ratio systems is the study of those of them
which undergo total eclipses.

\begin{figure}
\centerline{\includegraphics[width=6cm]{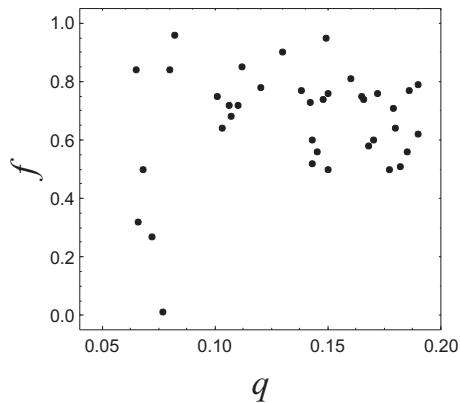}}
\caption{Relation $f-q$ for binaries with $q <$ 0.19.
\label{Fig4}}
\end{figure}

The analysis of the information in Table 3 led to the following
conclusions.

(1) The binaries with extreme low mass ratios are with periods
0.29--0.45 d. This result means that probably there are two types
progenitors of the merger: ultrashort-period binaries and binaries
with extreme small mass ratios.

(2) The components of binaries with extreme low mass ratios are of
intermediate spectral types with 6000 K $< T <$ 9800 K, i.e. there
are no late-type stars among them (in contrast to the
ultrashort-period binaries which are very late stars). Moreover,
there is a trend the target temperature to decrease with mass
ratio increase (Table 3).

(3) The extreme low mass ratios (Table 3) as a rule are
accompanied with big fill-out factors ($f >$ 0.27). However, our
target has almost contact configuration ($f\sim$ 0.014) and
presents serious exception.

The anonymous referee attracted our attention to the very
recent work of \cite{Liu+etal+18} who suppose a new mechanism of
long-term period variations for W UMa-type contact binaries.
Undergoing this mechanism, $f$ is oscillating. That means contact
binary systems with an extremely small $q$ can have very different
$f$, nearly from 0 to 1. This result can explain why the fill-out
factor of NSVS 2569022 could be so small.

(4) Some of binaries with extreme low mass ratios show long-term
sinusoidal modulation (\citealt{Sriram+etal+16}) probably due to
third body, but most of them reveal secular change of the orbital
periods, increasing or decreasing (Table 3). Theoretical and
evolutional study is necessary to explain which factor(s)
determine the two alternatives. The secular period variability may
be an indication of instability. Probably sometimes it is
connected with considerable mass loss via L2 (for deep-contact
configurations).

Besides observed by us eclipses we found only 3 light minima in V
filter during 2016 from AAVSO database and determined their times
(Table 4). These data cover two time ranges and did not allow to
search for a secular period changes of NSVS 2569022. We were able
only to improve its ephemeris
\begin{equation}
 HJD(MinI) = 2457068.47216(14) + 0.28779842(5) \times E .
\end{equation}

Besides the cases of extremely low mass ratio (Table 3), we found
studies of around thirty W UMa binaries with small mass ratios
($q<$ 0.19). We sampled their characteristics ($q, f, P, T$) in
Table 5. Most of the targets might be found also in the list of
\cite{Yang+Qian+15} but our sample does not contain systems with
photometric mass ratio of partially eclipsed targets due to their
poorly determined mass ratio. Table 5 presents also information
about the period variability of the binaries with $q<$ 0.19. It
turned out that almost all of them exhibit secular period
increase/decrease and some undergo quasiperiodic variation,
superimposed on a secular period change.

The trends (1--4) established for the binaries with the smallest
mass ratios (Table 3) are approximately valid for the targets with
mass ratio in the range 0.1--0.19 (Table 5). But there are several
exceptions with longer periods ($P >$ 0.45 d) and lower
temperatures ($T < $ 6000 K). Moreover, for the second sample
there is rather a trend the target temperature to increase with
the mass ratio increase (Table 5).

\cite{Yang+Qian+15} found empirical relation $f = 1.176 - 5.276
\times q + 11.64 \times q^2$ for their sample of low mass-ratio
deep-contact binaries. According to this relation NSVS 2569022
should have $f$=0.84.

The data from Table 3 and Table 5 reveal that the stars with mass
ratio within the range 0.10--0.19 have deep-contact configurations
(fill-out 0.5--0.9) while the binaries with extremely small mass
ratio within the narrow range 0.06--0.08 have fill-out factors in
the wide range 0.01--0.84 (Fig. 4). NSVS 2569022 is a record
holder with almost contact configuration.

\section{Conclusions}

The main results from the modeling of our photometric observations
of the W UMa binary NSVS 2569022 are: (i) The target undergoes
wide total eclipse and its mass ratio is determined with a big
confidence; (ii) The stellar components are of G spectral types;
(iii) NSVS 2569022 has slightly overcontact configuration (with
fill-out factor $f$ = 0.014); (iv) It is of W subtype; (v) The
mass ratio of only 0.077 ranks NSVS 2569022 in sixth place among
the binaries with smallest mass ratio. It implies that the target
probably goes to its instability and possible merger.

NSVS 2569022 is peculiar binary with an extremely small mass ratio
but slightly overcontact configuration.

\begin{acknowledgements}
The research was supported partly by project DN 08/20 of Scientific
Fundation of the Bulgarian Ministry of Education and Science as
well as by project RD-08-142 of Shumen University. The research
was made with the support of the private IRIDA Observatory
(www.irida-observatory.org).

This publication makes use of data products from the Two Micron
All Sky Survey, which is a joint project of the University of
Massachusetts and the Infrared Processing and Analysis
Center/California Institute of Technology, funded by the National
Aeronautics and Space Administration and the National Science
Foundation. This research also has made use of the SIMBAD
database, operated at CDS, Strasbourg, France, NASA Astrophysics
Data System Abstract Service. This paper makes use of data from
the DR7 of the WASP data (\citealt{Butters+etal+10}) as provided
by the WASP consortium, and the computing and storage facilities
at the CERIT Scientific Cloud, reg. no. $CZ.1.05/3.2.00/08.0144$
which is operated by Masaryk University, Czech Republic.
The authors are very grateful to the anonymous referee for
the valuable notes and recommendations.
\end{acknowledgements}


\label{lastpage}

\end{document}